\documentclass[prd,aps,showpacs,preprintnumbers,amsmath,amssymb,11pt]{revtex4}
\usepackage{graphicx}% Include figure files
\usepackage{dcolumn}% Align table columns on decimal point
\usepackage{bm}% bold math
\usepackage{psfrag}% psfrag

\begin{document}
%\draft  
\newcommand{\be}{\begin{equation}}\newcommand{\ee}{\end{equation}}
\newcommand{\bea}{\begin{eqnarray}}\newcommand{\eea}{\end{eqnarray}}
\newcommand{\bc}{\begin{center}}\newcommand{\ec}{\end{center}}
\def\no{\nonumber}
\def\eq#1{Eq. (\ref{#1})}\def\eqeq#1#2{Eqs. (\ref{#1}) and  (\ref{#2})}
%%%%%%%%%%%%%%%%%%%%%%%%%%%%%%%%%%%%%%%%%%%%%%%%%%%
\def\lsim{\raise0.3ex\hbox{$\;<$\kern-0.75em\raise-1.1ex\hbox{$\sim\;$}}}
\def\gsim{\raise0.3ex\hbox{$\;>$\kern-0.75em\raise-1.1ex\hbox{$\sim\;$}}}
\def\slash#1{\ooalign{\hfil/\hfil\crcr$#1$}}
\def\eff{\mbox{\tiny{eff}}}
\def\order#1{{\mathcal{O}}(#1)}
\def\pppm{B^0\to\pi^+\pi^-}
\def\pzpz{B^0\to\pi^0\pi^0}
\def\pppz{B^0\to\pi^+\pi^0}
%%%%%%%%%%%%%%%%%%%%%%%%%%%%%%%%%%%%%%%%%%%%%%%%%%%
%%\preprint{}
\title{$\eta-\eta^{\prime}$ mixing  } 
\author{
T. N. Pham}
\affiliation{
 Centre de Physique Th\'{e}orique, CNRS \\ 
Ecole Polytechnique, 91128 Palaiseau, Cedex, France }
\date{\today}
\begin{abstract}
The   $\eta -\eta^{\prime}$ mixing mass term 
due to the derivative coupling $SU(3)\times SU(3)$ symmetry breaking term,
produces  an additional momentum-dependent pole term for processes with
$\eta^{\prime}$, but is suppressed  in the  $\eta$ amplitude by a factor
$m_{\eta}^{2}/m_{\eta^{\prime}}^{2}$. This seems to be the 
origin of the two-angle description of the pseudo-scalar decay constants
used  in the literature. In this paper, by diagonalizing both the  mixing mass 
term and the momentum-dependent mixing term, we show that 
the $\eta -\eta^{\prime}$ system could be described by a meson field
renormalization and a new mixing angle $\theta$ which differs from the
usual mixing angle $\theta_{P}$ by a small momentum-dependent mixing $d$
term. This new mixing scheme with  exact treatment of the
momentum-dependent mixing term, is actually simpler than the perturbation
treatment and should be used in any determination of the $\eta
-\eta^{\prime}$  mixing angle and the 
momentum-dependent mixing term. Assuming 
nonet symmetry for the  $\eta_{0}$ singlet amplitude, from
the sum rules relating  $\theta$ and  $d$ to the measured vector meson
radiative decays amplitudes, we obtain consistent
solutions with  $\theta=-(13.99\pm 3.1)^{\circ}$,  $d=0.12\pm 0.03$ from $\rho\to\eta\gamma$  and  $\eta^{\prime}\to\rho\gamma$ decays, for  $\omega$ ,  $\theta=-(15.47\pm 3.1)^{\circ}$,  $d=0.11\pm 0.03$, and for
$\phi$,  $\theta=-(12.66\pm 2.1)^{\circ}$, $d=0.10\pm 0.03$. It seems
that vector meson radiative decays would favor a small
$\eta-\eta^{\prime}$ mixing angle and  a small momentum-dependent mixing  term. 

\end{abstract}
\maketitle
%%%%%%%%%%%%%%%%%%%%%%%%%%%%%%%%%%%%%%%%%%%%%%%%%%%

%%%%%%%%%%%%%%%%%%%%%%%%%%%%%%%%%%%%%%%%%%%%%%%%%%%

\section{Introduction}

The  $\eta-\eta^{\prime}$ mixing angle  used in the past to describe 
the $\eta-\eta^{\prime}$ system is based on the 
assumption that the off-diagonal octet-singlet mixing mass term does not
depend significantly on the energy of the state
\cite{Donoghue1}. However, as with the derivative coupling 
$SU(3)\times SU(3)$ breaking terms used in the derivation of the 
 $f_{K}/f_{\pi}$ ratio and the Callan-Treiman relation for the 
vector currents in $K_{l3}$ decays \cite{Pham1}, recent 
works\cite{Leutwyler,Kaiser} 
show that a quadratic derivative off-diagonal octet-singlet mixing term
could exist and requires two angles $\theta_{8}$ and $\theta_{0}$ to 
describe the pseudo-scalar meson decay constants. One could also describe the
 $\eta-\eta^{\prime}$ system by the  usual mixing angle $\theta_{P}$
 with  the additional off-diagonal derivative  $SU(3)$ breaking mass
 term   treated as a perturbation\cite{Pham2} in which  the momentum-dependent 
off-diagonal mass term produces  an additional contribution which is
suppressed by $O(m_{\eta}^{2}/m_{\eta^{\prime}}^{2})$ for  processes
involving $\eta$. Thus the quadratic momentum-dependent off-diagonal
mixing mass term, while  leaves the amplitude with $\eta$ almost
unaffected, could enhance or  suppress the $\eta^{\prime}$
amplitude. Since  mixing angle  contains higher order $SU(3)$ breaking
term, to  be  consistent, we need to include  also higher order terms in the  
momentum-dependent mixing  terms by diagonalizing  both the 
momentum-independent and  momentum-dependent mixing terms. In the 
past 20 years there have been  only two papers considering
diagonalizing  the $\eta -\eta^{\prime}$ Lagrangian with
both the  off-diagonal mass term and the full off-diagonal kinetic 
terms \cite{Schechter,Gedalin} which however produces an 
$\eta -\eta^{\prime}$ Lagrangian with  two  mixing angles and two
field renormalization  parameters. Actually, it is
not necessary to use their full off-diagonal kinetic terms, since the 
coefficients of the  $\partial_{\mu}\eta_{8}\,\partial_{\mu}\eta_{8}$ 
and $\partial_{\mu}\eta_{0}\,\partial_{\mu}\eta_{0}$ terms can be
absorbed into the mass terms after rescaling, so that the 
$\eta -\eta^{\prime}$ Lagrangian contains the usual canonical kinetic
 terms and only one  off-diagonal
 $\partial_{\mu}\eta_{8}\,\partial_{\mu}\eta_{0} $ term. With this most
 general kinetic  term, in this paper, we  will show that
the $\eta-\eta^{\prime}$ system could be described by 
the a new mixing angle and the renormalization of the $\eta$ and
$\eta^{\prime}$ meson fields. The new mixing angle contains the usual
mixing angle and a small  additional term coming from $d$.
This new mixing scheme with exact treatment of the momentum-dependent mixing
terms is actually simpler than the perturbation treatment in
\cite{Pham2}  and  should be used in any  determination of the
$\eta -\eta^{\prime}$  mixing angle and the  momentum-dependent mixing
term. In  this paper,  we  shall apply this new mixing
scheme to vector meson  radiative  decays . From the  sum  rules
relating the pure octet and
singlet vector meson radiative decay amplitudes to that for the
measured decays amplitudes, and using  nonet symmetry  for the pure octet 
and singlet amplitudes, we obtain  consistent solutions for the new 
mixing angle $\theta$ and the  momentum-dependent mixing term $d$. 
 For $\rho\to\eta\gamma$  and $\eta^{\prime}\to\rho\gamma$ decays, 
$\theta=-(13.99\pm 3.1)^{\circ}$,  $d=0.12\pm 0.03$, for
$\omega\to\eta\gamma$  and $\eta^{\prime}\to\omega\gamma$ decays,
$\theta=-(15.47\pm 3.1)^{\circ}$, $d=0.11\pm 0.03$ and for
$\phi\to\eta\gamma$. It is remarkable that
these values are consistent with each other, to within experimental
errors. For $\phi\to\eta^{\prime}\gamma$ decays, with $SU(3)$ breaking
from the $s$ quark magnetic coupling included, we get 
$\theta=-(12.66\pm 2.1)^{\circ}$, $d=0.10\pm 0.03$, consistent with the 
values for $\rho$ and $\omega$. After subtracting the $d$ terms, one 
would get a value of $-(8-10)^{\circ}$  for the usual mixing angle. It 
seems  that vector meson radiative decays would favor a small
 $\eta -\eta^{\prime}$ mixing angle as found in previous analysis, for
 example,  a value between
 $-13^{\circ}$ and $-17^{\circ}$, or an average $\theta_{P}= -15.3^{\circ} \pm 1.3^{\circ}$ is obtained \cite{Bramon} and $\theta_{P}\approx -11^{\circ}$
is obtained in \cite{Benayoun}, also a  recent analysis
\cite{Escribano,KLOE} using the more  precise $V\to P\gamma$ measured
branching ratios \cite{PDG} found  $\theta_{P}=-13.3^{\circ} \pm
1.3^{\circ}$. Our values for $d$ is also smaller than the chiral perturbation
values and other phenomenological analysis in the two-angle mixing
approach\cite{Kaiser,Feldmann,Chen}. In the next section we 
will obtain the diagonalized  Lagrangian for the $\eta-\eta^{\prime}$
 system with the new $\eta-\eta^{\prime}$ mixing angle $\theta$.

\section{The diagonalized $\eta-\eta^{\prime}$ Lagrangian}

 We begin by  writing down the Lagrangian for the $\eta-\eta^{\prime}$
system with the usual non-derivative mixing mass term $m_{08}^{2}$,
the pure octet $\eta_{8}$ mass $m_{8}^{2}$,  the  singlet  $\eta_{0}$ 
mass $m_{0}^{2}$, and the derivative $\eta_{0}-\eta_{8} $ mixing term.

\be
{\cal L}_{0} = \frac{1}{2}(\partial_{\mu}\eta_{8}\,\partial_{\mu}\eta_{8} +
\partial_{\mu}\eta_{0}\,\partial_{\mu}\eta_{0} + m_{8}^{2}\eta_{8}^{2}+
 m_{0}^{2}\eta_{0}^{2}) +  d\,\partial_{\mu}\eta_{8}\,\partial_{\mu}\eta_{0} + m_{08}^{2}\eta_{8}\eta_{0}
\label{L0}
\ee
where  the strength $d$ is given by $L_{5}$ and  higher 
order terms in Chiral Perturbation theory
\cite{Leutwyler,Kaiser,Gasser}. The $\eta_{0}-\eta_{8} $ Lagrangian in
Eq. (\ref{L0}) contains the most general kinetic and mass term. The full 
off-diagonal kinetic and mass terms used in previous 
works \cite{Schechter,Gedalin} to diagonalize both the  kinetic and mass 
terms of the  $\eta_{0}-\eta_{8} $ system,  can be brought to the 
above  form since the rescaling of the kinetic terms can be
absorbed into the mass term so that ${\cal L}_{0} $  contains only
the off-diagonal $\partial_{\mu}\eta_{8}\,\partial_{\mu}\eta_{0} $
and the usual  canonical kinetic terms. Thus our Lagrangian contains,
as mentioned earlier, besides the usual $\eta -\eta^{\prime}$ mass
parameters, only two mixing  parameters, the usual momentum-independent
$\eta -\eta^{\prime}$ mixing  mass term and the momentum-dependent
 $\eta -\eta^{\prime}$ off-diagonal
kinetic terms. This is an  important difference   between our  approach 
and that of Ref. \cite{Schechter,Gedalin}. In a straightforward manner, 
we will show that   the $\eta -\eta^{\prime}$ system can be described by 
only one mixing angle and a field renormalization parameter.

 To  diagonalize this  Lagrangian, we shall first make the substitution~:
\be
\eta_{8}= \frac{(\eta_{01}-\eta_{81})}{\sqrt{2}}, \quad
\eta_{0}= \frac{(\eta_{01}+\eta_{81})}{\sqrt{2}}.
\ee
${\cal L}_{0} $ becomes,
\be
{\cal L}_{1} = \frac{(1-d)}{2}\partial_{\mu}\eta_{81}\,\partial_{\mu}\eta_{81} +
\frac{(1+d)}{2}\partial_{\mu}\eta_{01}\,\partial_{\mu}\eta_{01} + \frac{1}{2}(m_{81}^{2}\eta_{81}^{2}+ m_{01}^{2}\eta_{01}^{2})   + m_{081}^{2}\eta_{81}\eta_{01}
\label{L1}
\ee
with
\be
m_{81}^{2}= \frac{(m_{0}^{2}+m_{8}^{2}-2m_{08}^{2})}{2}, \quad
m_{01}^{2}= \frac{(m_{0}^{2}+m_{8}^{2}+2m_{08}^{2})}{2}, \quad
m_{081}^{2}= \frac{(m_{0}^{2}-m_{8}^{2})}{2}.
\ee
To bring the kinetic term in ${\cal L}_{1} $ to the canonical form, we now perform a 
renormalization of $\eta_{81} $ and $\eta_{01} $ meson field operators:
\be   
 \eta_{81} =\frac{\eta_{82}}{\sqrt{1-d}},\quad
 \eta_{01} =\frac{\eta_{02}}{\sqrt{1+d}}
\ee
and ${\cal L}_{1} $ becomes  
\be
{\cal L}_{2} = \frac{1}{2}\biggl(\partial_{\mu}\eta_{82}\,\partial_{\mu}\eta_{82} +
\partial_{\mu}\eta_{02}\,\partial_{\mu}\eta_{02} + \frac{m_{81}^{2}}{(1-d)}\eta_{82}^{2}+
 \frac{m_{01}^{2}}{(1+d)}\eta_{02}^{2}\biggr) +  \frac{m_{081}^{2}}{\sqrt{1-d^{2}}}\eta_{82}\eta_{02}
\label{L2}
\ee
which can now be brought back to  the octet-singlet basis by
the transformation:
\be
\eta_{82}= \frac{(\eta_{03}-\eta_{83})}{\sqrt{2}},  \quad
\eta_{02}= \frac{(\eta_{03}+\eta_{83})}{\sqrt{2}}.
\ee
We have finally,
\be
{\cal L}_{3} = \frac{1}{2}(\partial_{\mu}\eta_{83}\,\partial_{\mu}\eta_{83} +
\partial_{\mu}\eta_{03}\,\partial_{\mu}\eta_{03} + m_{82}^{2}\eta_{83}^{2}+
 m_{02}^{2}\eta_{03}^{2}) +  m_{082}^{2}\eta_{83}\eta_{03}
\label{L3}
\ee
with 
\bea
&& m_{82}^{2}= \frac{(1-\sqrt{1-d^{2}})m_{0}^{2}+(1+\sqrt{1-d^{2}})m_{8}^{2}}{2(1-d^{2})}+ \frac{d\,m_{08}^{2}}{(1-d^{2})}, \nonumber\\
&& m_{02}^{2}= \frac{(1+\sqrt{1-d^{2}})m_{0}^{2}+(1-\sqrt{1-d^{2}})m_{8}^{2}}{2(1-d^{2})}+ \frac{d\,m_{08}^{2}}{(1-d^{2})}, \nonumber\\
&& m_{082}^{2}= \frac{m_{08}^{2} -d(m_{0}^{2}+m_{8}^{2})/2}{(1-d^{2})}.\label{cP}
\eea
 Thus we have been able  to bring the original Lagrangian of the
pure octet $\eta_{8}$ and singlet $\eta_{0}$ mesons with 
the derivative coupling $SU(3)$ symmetry breaking momentum-dependent
 $\eta_{8}-\eta_{0}$ mixing term, to the usual form with only the 
energy-independent mixing mass term with ${\cal L}_{3} $  having
the same form as ${\cal L}_{0} $, except that the mass  and mixing
terms are modified by  additional contributions from the momentum-dependent 
mixing term $d$ and the renormalization of the $\eta_{8}$ and $\eta_{0}$
meson fields, and in the limit of $d=0$, we recover the usual mass term in
${\cal L}_{0} $. In terms of $\eta_{83}$ and $\eta_{03}$ state, the
pure $SU(3)$ octet and singlet state are then given by
\bea
&&\eta_{8}= \biggl(\frac{\sqrt{1-d} +\sqrt{1+d}}{2\sqrt{(1-d^{2})}}\biggr)\eta_{83} + 
\biggl(\frac{\sqrt{1-d} -\sqrt{1+d}}{2\sqrt{(1-d^{2})}}\biggr)\eta_{03} ,\nonumber \\
&&\eta_{0}= \biggl(\frac{\sqrt{1-d} -\sqrt{1+d}}{2\sqrt{(1-d^{2})}}\biggr)\eta_{83} + 
\biggl(\frac{\sqrt{1-d} +\sqrt{1+d}}{2\sqrt{(1-d^{2})}}\biggr)\eta_{03} .
\label{eta08}
\eea
 From the above expressions, we see that $\eta_{83}$ and $\eta_{03}$
 states  are mixture of the pure $\eta_{8}$ and $\eta_{0}$ and 
becomes the pure octet and singlet state in the limit
of $d=0$. This is an example of mixing caused by renormalization of 
the field operators due to the momentum-dependent derivative coupling 
$SU(3)$ breaking terms. The Lagrangian in Eq. (\ref{L3} ) can now be
brought to the diagonal form by writing $\eta_{83}$ and $\eta_{03}$
in terms of the physical $\eta$ and $\eta^{\prime}$ states and the mixing
angle $\theta$~:
\bea
&&\eta_{83}= \cos(\theta)\eta  +\sin(\theta)\eta^{\prime} ,\nonumber \\
&&\eta_{03}= -\sin(\theta)\eta +\cos(\theta)\eta^{\prime} .
\label{eta-etap}
\eea
with $\theta$ given by:
\be
\tan(2\,\theta) = \frac{2\,m_{08}^{2}-d\,(m_{0}^{2}+m_{8}^{2})}{(m_{0}^{2}-m_{8}^{2})\sqrt{1-d^{2}}}
\label{theta}
\ee
or
\be
\sin(\theta) = \biggl(\frac{\cos(2\,\theta)}{\cos(\theta)}\biggr)\biggl(\frac{m_{08}^{2}-d\,(m_{0}^{2}+m_{8}^{2})/2}{(m_{0}^{2}-m_{8}^{2})\sqrt{1-d^{2}}}\biggr)
\label{sintheta}
\ee
which takes a simple form for $\theta$ small,
\be
\sin(\theta) = \biggl(\frac{m_{08}^{2}-d\,(m_{0}^{2}+m_{8}^{2})/2}{(m_{0}^{2}-m_{8}^{2})\sqrt{1-d^{2}}}\biggr)
\label{sintheta0}
\ee
After this  last step, we arrive at the Lagrangian:
\be
{\cal L} =\frac{1}{2}(\partial_{\mu}\eta\,\partial_{\mu}\eta +\partial_{\mu}\eta^{\prime}\,\partial_{\mu}\eta^{\prime} + {m_{\eta}}^{2}\eta^{2}+
 {m_{\eta^{\prime}}}^{2}{\eta^{\prime}}^{2}) 
\label{L}
\ee
with  ${m_{\eta}}^{2} $ and   ${m_{\eta^{\prime}}}^{2}$ given by:
\bea
&&{m_{\eta}}^{2}= \frac{(m_{0}^{2} +
  m_{8}^{2}-2\,d\,m_{08}^{2})}{2(1-d^{2})} - \frac{(m_{0}^{2} -
  m_{8}^{2})\cos(2\theta)}{2\sqrt(1-d^{2})} +\frac{( d\,(m_{0}^{2} + m_{8}^{2})-2\,m_{08}^{2})\,\sin(2\theta) }{2\,(1 -d^{2})}    \nonumber\\    
&&{m_{\eta^{\prime}}}^{2}= \frac{(m_{0}^{2} +
  m_{8}^{2}-2\,d\,m_{08}^{2})}{2(1-d^{2})} + \frac{(m_{0}^{2} -
  m_{8}^{2})\cos(2\theta)}{2\sqrt(1-d^{2})} -\frac{( d\,(m_{0}^{2} + m_{8}^{2})-2\,m_{08}^{2})\,\sin(2\theta) }{2\,(1 -d^{2})}           
\label{meta00}
\eea
 The pure octet  $\eta_{8}$ and singlet $\eta_{0}$  can now be expressed
terms of  $\eta$ and $\eta^{\prime}$. From 
 Eqs. (\ref{eta08}-\ref{eta-etap}), we have~: 
\be
\eta_{8}= C_{8\eta}\,\eta + C_{8\eta^{\prime}}\,\eta^{\prime}, \quad \quad
\eta_{0}= C_{0\eta}\,\eta + C_{0\eta^{\prime}}\,\eta^{\prime}.
\label{eta08p}
\ee
with 
\bea
&&C_{8\eta}=\biggl(-\frac{(\sqrt{1-d}  -\sqrt{1+d})\sin(\theta)}{2\sqrt{(1-d^{2})}}
+ \frac{(\sqrt{1-d}  +\sqrt{1+d})\cos(\theta)}{2\sqrt{(1-d^{2})}}\biggr)\nonumber\\
&&C_{8\eta^{\prime}}=\biggl(\frac{(\sqrt{1-d} -\sqrt{1+d})\cos(\theta)}{2\sqrt{(1-d^{2})}}
+ \frac{(\sqrt{1-d}  +\sqrt{1+d})\sin(\theta)}{2\sqrt{(1-d^{2})}}\biggr)\nonumber\\
&&C_{0\eta}=\biggl(-\frac{(\sqrt{1-d}  +\sqrt{1+d})\sin(\theta)}{2\sqrt{(1-d^{2})}}
+ \frac{(\sqrt{1-d}  -\sqrt{1+d})\cos(\theta)}{2(1-d^{2})}\biggr)
\nonumber\\
&&C_{0\eta^{\prime}}=\biggl(\frac{(\sqrt{1-d}  +\sqrt{1+d})\cos(\theta)}{2\sqrt{(1-d^{2})}}
+ \frac{(\sqrt{1-d}  -\sqrt{1+d})\sin(\theta)}{2\sqrt{(1-d^{2})}}\biggr)
\label{Ceta}
\eea
For $d=0$, we recover the usual expression given in Eq. (\ref{eta-etap}) .

To first order in $d$, we have,
\bea
&&\eta_{8}= \biggl(d\sin(\theta)/2+ \cos(\theta)\biggr)\eta+
\biggl(-d\cos(\theta)/2 + \sin(\theta)\biggr)\eta^{\prime}  ,\nonumber\\
&&\eta_{0}= \biggl(-\sin(\theta)-d\cos(\theta)/2\biggr)\eta +\biggl(\cos(\theta) -d\sin(\theta)/2\biggr)\eta^{\prime} 
\label{eta08p0}
\eea

Consider now the $d$ terms in Eq. (\ref{eta08p0}). The
contribution to $\eta^{\prime}$ amplitude from the pure $\eta_{8}$ 
term is proportional to  $(-d\cos(\theta)/2+ \sin(\theta)) $
which gives $-d/2$ from the first term, while another $-d/2$
from the $\sin(\theta)$ term. Similarly, the $d$ term 
in  the $\eta$ amplitude coming from the pure singlet $\eta_{0}$ term
$(\sin(\theta)+ d\,\cos(\theta)/2) $ cancels out ( $\sin(\theta)$
having  the same  $d$ term with opposite sign). More precisely, to first order
in $d$, and  neglecting also $\sin(\theta/2)^{2}$ term in
$\cos(\theta)$, we have from Eq. (\ref{sintheta0}):
\bea
&&\eta_{8}= \biggl(d\sin(\theta)/2+ \cos(\theta)\biggr)\eta+
 \biggl(\sin(\theta_{P})+\frac{d\,m_{0}^{2}}{(m_{0}^{2}-m_{8}^{2})}\biggr)\eta^{\prime} 
 ,\nonumber\\
&&\eta_{0}= \biggl(-\sin(\theta_{P})+\frac{d\,m_{8}^{2}}{(m_{0}^{2}-m_{8}^{2})}
\biggr)\eta +\biggl(\cos(\theta)-d\sin(\theta)/2\biggr)\eta^{\prime} 
\label{eta08p0f}
\eea
where $\theta_{P}$ is the mixing angle for $d=0$ (the usual mixing angle).

 This agrees with the perturbation treatment of the derivative 
$SU(3)\times SU(3)$ symmetry breaking terms given in \cite{Pham2},
except for the $d\sin(\theta)$ term which is second order in $SU(3)$ breaking.

 We see that in the presence of the momentum-dependent mixing term $d$,
the $\eta$ and $\eta^{\prime}$ amplitude now depend on both $\theta$ and
$d$ and are given completely by Eqs. (\ref{Ceta}). Obviously this
simple expressions should be used in
any physical processes with  $\eta$ and $\eta^{\prime}$ rather than 
the perturbation treatment of the momentum-dependent mixing term used
in \cite{Pham2}. Given  $A_{8}, A_{0}$,  the octet and  singlet
amplitude for  $\eta_{8}$ and $\eta_{0}$, respectively,  the physical 
amplitudes are then~:
\bea
&&A_{\eta}=C_{8\eta}\,A_{8} + C_{0\eta}\,A_{0}\nonumber\\
&&A_{\eta^{\prime}}= C_{8\eta^{\prime}}\,A_{8} + C_{0\eta^{\prime}}\,A_{0}
\label{A08}
\eea
 
Following the two-angle mixing approach\cite{Kaiser}, consider now
the quantity
\be
P_{08}=A_{8}A_{0}(C_{8\eta}\,C_{0\eta} + C_{8\eta^{\prime}}C_{0\eta^{\prime}})
\label{P08}
\ee
using Eq. (\ref{Ceta}), we find
\be
P_{08}= -A_{8}A_{0}\frac{d}{(1-d^{2})}= -A_{8}A_{0}\sin(\theta_{0}-\theta_{8})
\label{P08v}
\ee
which is precisely the expression obtained in Chiral Perturbation Theory.
To first order in $d$, $\sin(\theta_{0}-\theta_{8})=d$.
This  shows clearly that the parameter $d$ is directly proportional to 
the coefficient $L_{A}$ in the derivative expansion\cite{Kaiser}.
For $d=0$, $P_{08}=0$, we recover the orthogonality of the unitarity
transformation between  physical and unmixed states with the usual mixing 
angle.

   Using  Eq. (\ref{theta}) to express $m_{08}^{2}$ in terms of
 $\tan(2\theta) $, the  expressions for $\eta$ and $\eta^{\prime}$ masses in 
Eq. (\ref{meta00}) are then:
\bea
&&{m_{\eta}}^{2}=  \frac{(m_{0}^{2} + m_{8}^{2})}{2} - \frac{(m_{0}^{2} - m_{8}^{2})}{2\sqrt(1-d^{2})\cos(2\theta)}-\frac{(d\,\tan(2\theta))(m_{0}^{2} - m_{8}^{2}) }{2\sqrt(1 -d^{2})}  \nonumber\\
&&{m_{\eta^{\prime}}}^{2}= \frac{(m_{0}^{2} + m_{8}^{2})}{2} + \frac{(m_{0}^{2} - m_{8}^{2})}{2\sqrt(1-d^{2})\cos(2\theta)} -\frac{( d\,\tan(2\theta))(m_{0}^{2} - m_{8}^{2}) }{2\sqrt(1 -d^{2})}  
\label{meta1}
\eea
which now depend only on $m_{0}^{2}$, $m_{8}^{2}$ and $d$. By taking
the mass difference ${m_{\eta^{\prime}}}^{2}- m_{8}^{2}$ and
${m_{\eta}}^{2} - m_{8}^{2}$,  we obtain:
\bea
&&{m_{\eta}}^{2}-m_{8}^{2}=  \frac{(m_{0}^{2} - m_{8}^{2})}{2}\biggl(1 - \frac{1}{\sqrt(1-d^{2})\cos(2\theta)}-\frac{d\,\tan(2\theta)}{\sqrt(1 -d^{2})} \biggr)\nonumber\\
&&{m_{\eta^{\prime}}}^{2}-m_{8}^{2}= \frac{(m_{0}^{2} - m_{8}^{2})}{2}\biggl(1 + \frac{1}{\sqrt(1-d^{2})\cos(2\theta)} -\frac{ d\,\tan(2\theta)}{\sqrt(1 -d^{2})}  \biggr)      
\label{meta2}
\eea
This implies,
\be
{m_{\eta}}^{2}-m_{8}^{2} = R\,({m_{\eta^{\prime}}}^{2}-m_{8}^{2}).
\label{Reta}
\ee
with $R$ given by:
\be
R= -\biggl(1 -\sqrt(1-d^{2})\cos(2\theta)  +d\,\sin(2\theta)
\biggr)\biggl(1+ \sqrt(1-d^{2})\cos(2\theta)  -d\,\sin(2\theta)\biggr)^{-1}
\label{R}
\ee
As $d$ is a small $SU(3)\times SU(3)$ breaking parameter, putting 
 $d=\sin(\alpha)$ and $\sqrt{1-d^{2}}= \cos(\alpha)$, the above
 expression Eq. (\ref{R}) takes a simple form,
\be
R= -\tan(\theta+\alpha/2)^{2}
\label{Rs}
\ee
For small $d$, $\alpha\approx \sin(\alpha)=d$, $\theta +\alpha/2 \approx
\theta_{P} $,  and $R$ is essentially the usual relation 
$R=-\tan(\theta_{P})^{2}$ which is not affected by the presence of a 
momentum-dependent mixing term.

\section{Mixing angle from vector meson radiative decays}

 With  our Lagrangian in the diagonal form, we shall now try to  determine
$\theta$ and $d$ using the sum rules \cite{Pham2}, obtained by equating
 the vector meson radiative decay matrix element for the pure 
octet $\eta_{8}$ and singlet $\eta_{0}$ with the expressions
for these quantities extracted from the measured matrix elements with
 $\eta$ and $\eta^{\prime}$ given by Eq. (\ref{eta08p}). Defining, as
 in \cite{Pham2}, the electromagnetic form factor $V\to P$ by:
\be
<P(p_{P})|J^{\rm em}_{\mu}|V(p_{V})> = \epsilon_{\mu
  p_{P}p_{V}\epsilon_{V}}g_{VP\gamma}
\label{fVP}
\ee
where  $g_{VP\gamma} $ is the on-shell $VP\gamma$ coupling constant
with  dimension  the inverse of energy. We have, for the radiative decay
rates \cite{Ball}
\bea
&&\Gamma(V\to P\gamma)= \frac{\alpha}{24}g^{2}_{VP\gamma}
\Biggl(\frac{m_{V}^{2}-m_{P}^{2}}{m_{V}}\biggr)^{3}\nonumber\\
&&\Gamma(P\to V\gamma)= \frac{\alpha}{8}g^{2}_{VP\gamma}
\Biggl(\frac{m_{P}^{2}-m_{V}^{2}}{m_{P}}\biggr)^{3}
\label{VPrate}
\eea

\begin{table}[ht]
\begin{tabular}{|c|c|c|c|}
\hline
 Decay &$g_{VP\gamma}$,$P=\eta_{8},\eta_{0}$&$g_{VP\gamma}(\rm
 exp.)$&\rm BR(exp) \cite{PDG} \\ 
\hline
\hline
$\rho^{\pm}\to \pi^{\pm}\gamma$&$(1/3)\,g_{u}$ &$0.72\pm 0.04$ &$ (4.5\pm 0.5)\times 10^{-4}$\\
$\rho^{0}\to \pi^{0}\gamma$&$(1/3)\,g_{u}$  &$0.83\pm 0.05$ & $ (6.0\pm 0.8)\times 10^{-4}$\\
$\rho^{0}\to \eta\gamma$&$0.58\,g_{u}\,(f_{\pi}/f_{\eta_{0}})$ &$1.59\pm 0.06$&$(3.00\pm 0.20)\times 10^{-4}$ \\
$ \omega\to \pi^{0}\gamma$&$0.99\,g_{u}$&$2.29\pm 0.03$ &$(8.28\pm 0.28)\%$ \\
$  \omega\to \eta\gamma$&$0.17\,g_{u}\, (f_{\pi}/f_{\eta_{0}})$&$0.45 \pm 0.02$  &$(4.6\pm 0.4)\times 10^{-4} $\\
 $  \phi\to \pi^{0}\gamma$&$0.06\,g_{u}$ &$0.13\pm 0.003$&$(1.27\pm 0.06)\times 10^{-3}$\\
$ \phi\to \eta\gamma$ &$0.47\,g_{u}\, (f_{\pi}/f_{\eta_{0}})$ &$0.71\pm 0.01 $&$(1.309\pm 0.024)\%$\\
$ \phi\to \eta^{\prime}\gamma$ &$-0.31\,g_{u}\, (f_{\pi}/f_{\eta_{0}}) $ &$-(0.72\pm 0.01) $&$(6.25\pm 0.21)\times 10^{-5}$\\
$ \eta^{\prime}\to \rho^{0}\gamma$ &$0.82\,g_{u}\, (f_{\pi}/f_{\eta_{0}}) $ &$1.35\pm 0.02 $&$(29.1\pm 0.5)\%$\\
$ \eta^{\prime}\to \omega\gamma$ &$0.29\,g_{u}\, (f_{\pi}/f_{\eta_{0}}) $ &$0.44\pm 0.02 $&$(2.75\pm 0.23)\%$\\
$K^{*\pm}\to K^{\pm}\gamma$&$0.38\,g_{u}\, (f_{\pi}/f_{K})$ &$0.84\pm 0.04$ &$ (9.9\pm 0.9)\times 10^{-4}$\\
$K^{*0}\to K^{0}\gamma$&$-0.62\,g_{u}\, (f_{\pi}/f_{K})$  &$-(1.27\pm 0.05)$ & $ (2.46\pm 0.22)\times 10^{-3}$\\
\hline
\end{tabular}
\caption{ Theoretical values for $V\to P\gamma$ with $P=\eta_{8},\eta_{0}$
together with the measured branching ratios and the extracted
$g_{VP\gamma}$, taken from Ref. \cite{Pham2}}
\label{tab-1}
\end{table}
  For convenience, we give in  Table. \ref{tab-1},  the measured radiative
branching ratios together with the extracted coupling
constant $g_{VP\gamma}$ in  unit of $\rm GeV^{-1}$ and its theoretical value 
derived  either from an $SU(3)$ effective Lagrangian with nonet
symmetry for the $V\to\eta_{0}\gamma$ amplitude or from the 
quark counting rule with the coupling constant
$g_{VP\gamma}$ given in terms of the quark coupling constant  $g_{q}$,
($q=u,d,s$) for the magnetic  transition 
$(q\bar{q})(1^{-})\to (q\bar{q})(0^{-})\gamma$ \cite{Bramon,Escribano,Ball}.
More details on the theoretical values for $V\to \eta_{8}\gamma$ and
 $V\to \eta_{0}\gamma$ can be found in Ref. \cite{Pham2}.

In  terms of $g_{VP\gamma}$, the sum rules read:
\bea
&&S(V\to\eta\gamma)=g_{V\eta\gamma}\,C_{8\eta} +
g_{V\eta^{\prime}\gamma}\,C_{8\eta^{\prime}}=  \biggl(\frac{g_{V\eta_{8}\gamma}}{g_{V\pi^{0}\gamma}}\biggr)_{\rm th.}g_{V\pi^{0}\gamma}
\nonumber\\
&&S(\eta^{\prime}\to V\gamma)= g_{V\eta\gamma}\,C_{0\eta} +
g_{V\eta^{\prime}\gamma}\,C_{0\eta^{\prime}}=  \biggl(\frac{g_{V\eta_{0}\gamma}}{g_{V\pi^{0}\gamma}}\biggr)_{\rm th.}g_{V\pi^{0}\gamma}
\label{sum}
\eea
and similarly for other vector meson radiative decays.
Thus with the updated values of the measured values for $g_{VP\gamma}$
 in Table. \ref{tab-1}, we have, for $\rho$ meson radiative decay:
\bea
&&S(\rho\to\eta\gamma)=1.59\,C_{8\eta} + 1.35\,C_{8\eta^{\prime}}=1.12
\nonumber\\
&&S(\eta^{\prime}\to\rho\gamma)=1.59\,C_{0\eta} +1.35\,C_{0\eta^{\prime}}=1.63
\label{rho}
\eea
and for $\omega$ meson~:
\bea
&&S(\omega\to\eta\gamma)=0.45\,C_{8\eta} + 0.44\,C_{8\eta^{\prime}}=0.29
\nonumber\\
&&S(\eta^{\prime}\to\omega\gamma)=0.45\,C_{0\eta} +0.44\,C_{0\eta^{\prime}}=0.53
\label{omega}
\eea
 From the sets of the above equations, we obtain the following solutions 
for $\theta$ and $d$:
\bea
&&\theta=-(13.99\pm 3.1)^{\circ}, \quad   d=0.12\pm 0.03 ,\quad 
\mbox{for $\rho$}\nonumber\\
&&\theta=-(15.48\pm 3.1)^{\circ}, \quad   d=0.11\pm 0.03 , \quad 
\mbox{for $\omega$}
\eea
Since $\eta-\eta^{\prime}$ mixing is an $SU(3)$ breaking effect not
present in the $\eta_{8}$ and $\eta_{0}$ decay amplitudes, $\rho$ meson
radiative decays in which only $u,d$ quark are active,  offer a rare
opportunity to determine the mixing angle  free from uncertainties
from  $SU(3)$ breaking due to $s$ quark magnetic coupling which is present
in radiative $\phi$ meson radiative decays. With an almost ideal mixing, 
$\omega$ meson  radiative decays is also rather insensitive to  the 
$s$ quark magnetic coupling $SU(3)$ breaking, which is rather small, of 
the order of $1.5\%$. In fact, as shown in Ref.\cite{Pham2},  instead
of $\omega$ radiative decay amplitudes  alone, one can use a linear 
combination for an ideal  mixing state,  the $\omega_{0}$ state with 
the decay amplitudes with  only $u,d$  quark active . We have 
($\varphi_{V}=(3.2\pm 0.1)^{\circ}$), 
\be
S(\omega_{0}\to\eta\gamma)= \cos\varphi_{V}\,S(\omega\to\eta\gamma)+
\sin\varphi_{V}\,S(\phi\to\eta\gamma)\label{sot}
\ee
and  similar expression for $S(\eta^{\prime}\to \omega_{0}\gamma)$. The
solutions for this ideal mixing case is then,
\be
\theta=-(15.40\pm 2.1)^{\circ}, \quad  d=0.12\pm 0.03 ,  \quad\mbox{for $\omega_{0}$}
\label{omega0}
\ee
which is very close to the solution for $\omega$. This indicates that
$SU(3)$ breaking due to $s$ quark magnetic coupling in $\omega$ radiative
decay   is indeed quite small and can be neglected. This ideal mixing solution
is also consistent with the solution  for $\rho$. Taking the average of
the solution for $\rho$ and $\omega_{0}$, we have:
\be
\theta=-(14.68\pm 3.1)^{\circ}, \quad   d=0.115\pm 0.03,
\label{average}
\ee
 For $\phi$ meson, from the sum rules
\bea
&&S(\phi\to\eta\gamma)=0.71\,C_{8\eta} - 0.72\,C_{8\eta^{\prime}}=0.88
\nonumber\\
&&S(\phi\to\eta^{\prime}\gamma)=0.71\,C_{0\eta} -0.72\,C_{0\eta^{\prime}}=-0.59
\label{phi}
\eea
the solution is then
\be
\theta=-(12.66\pm 2.1)^{\circ}, \quad  d=0.10\pm 0.03 ,  \quad
\mbox{for $\phi$}
\label{ss2}
\ee
consistent with the corresponding values for $\rho$ and $\omega$ given
by  Eq. (\ref{omega}) and  Eq. (\ref{average}). This indicates
that $SU(3)$ breaking for $\phi$  meson radiative decays  is correctly
given by  $K^{*}\to K\gamma$ decays  for which the   new  measured
 branching ratio gives $k=0.83\pm 0.04$, close
to the value $k=0.85$ given above. Thus the values we obtained from
$\phi$ meson radiative decays are used as a way to check $SU(3)$
breaking effect in $\phi\to \eta\gamma,\eta^{\prime}\gamma$ decays  rather
than a determination of the mixing angle.  The value for
 $d$,($d=\sin(\theta_{0}-\theta_{8})$) obtained above with our diagonalized
Lagrangian is smaller than the values obtained in  chiral perturbation
and  other phenomenological analysis\cite{Kaiser,Feldmann,Chen} which give
$(\theta_{0}-\theta_{8})$ in the range $(12-17)^{\circ}$ in the 
 two-angle mixing approach.

 Thus, by treating exactly the derivative coupling mixing term
with our diagonalized Lagrangian  we have found a small mixing
angle in  vector meson radiative decays which are also found to have
small mixing  angle (the usual mixing angle)  in previous 
works\cite{Bramon,Benayoun,Escribano,KLOE}. By subtracting the $d$ term
in $\theta$, we obtain a value $-(8-10)^{\circ}$ for the usual 
mixing angle. This value is  smaller by a few  degrees than the 
values we obtained in our  previous work \cite{Pham2}. This could be 
due to the  exact treatment of the momentum-dependent mixing term in 
our Lagrangian.

\section{Conclusion}

 In conclusion, we have diagonalized both the mass term and the
momentum-dependent mixing term in the $\eta-\eta^{\prime}$
Lagrangian and shown that the $\eta-\eta^{\prime}$ system can be described
by two parameters, the meson field renormalization and a new 
$\eta-\eta^{\prime}$ mixing angle which differs from the usual mixing 
angle by a small momentum-dependent mixing term. The expressions
for $\eta$ and $\eta^{\prime}$ amplitude in our new mixing scheme are 
actually  quite simple and should be used for any process with
$\eta$ and $\eta^{\prime}$. Using  the measured  vector meson radiative
decays, we obtain consistent solutions for the  mixing angle and the 
momentum-dependent mixing  term. The small mixing angle we found is 
consistent with previous  determinations. It seems that  vector meson
radiative decays would favor  a small $\eta-\eta^{\prime}$ mixing angle
$\theta$  and a small momentum-dependent mixing  term $d$. 

%%%%%%%%%%%%%%%%%%%%%%%%%%%%%%%%%%%%%%%%%%%%%%%%%%%

\end{document}